\documentclass{iopart}
\usepackage{graphicx}
\usepackage{bm}
\usepackage{amssymb,bbm,epsfig,color}
\usepackage{iopams}

\newcommand{\text}[1]{{\rm #1}}
\newcommand{\operatorname}[1]{\text{#1}}

\begin{document}
{\vspace*{-3\baselineskip}\hspace*{\fill}\sf published in: J.~Phys.~A:~Math.~Theor. 42, 362001 (2009) }

\jl{1}

\title{Emergence of pointer states in a non-perturbative environment}

\author{Marc Busse and Klaus Hornberger}
\address{Arnold Sommerfeld Center for Theoretical Physics, 
Ludwig-Maximilians-Universit{\"a}t M{\"u}nchen,\\ Theresienstra{\ss}e 37, 
80333 Munich, Germany
}

\date{\today}

\begin{abstract}
We show that the pointer basis distinguished by collisional decoherence consists of exponentially localized, solitonic wave packets.
Based on the orthogonal unraveling of the quantum master equation,
we characterize their formation and dynamics, and we demonstrate that the statistical weights arising from an initial superposition state are given by the required projection. Since the spatial width of the pointer states can be obtained by accounting for the gas environment in a microscopically realistic fashion, one may thus calculate the coherence length of a strongly interacting gas.

\end{abstract}


%

\section{Introduction}

If a quantum system is brought into contact with
an environment, the states forming its Hilbert space are no longer equals
among each other. Rather, a particular set of wave functions, the {{\em
pointer states\/}}, is singled out by the environmental interaction
{\cite{ZurekPointer3mal}}. These states are distinguished by the fact that
they retain their purity for a relatively long time, say, under a Markovian
master equation describing the environmental influence, while superpositions
of different pointer states decay rapidly into a mixture. The associated
decoherence rate can be many orders of magnitude faster than any other
relevant time scale, including that of the dissipative effects induced by the
environmental coupling {\cite{Joos2003a,Schlosshauer2007a}}.

The concept of pointer states plays a fundamental role in explaining both the
emergence of classicality and the operation of quantum measurement devices
within the framework of quantum theory
{\cite{ZurekPointer3mal,Joos2003a,Schlosshauer2007a}}. In practice, knowing
the pointer states and their time evolution allows one to directly specify the
dynamics of an arbitrary initial state after the decoherence time, without
solving the master equation. It is thus important to understand the emergence
of pointer states in microscopically realistic environments, to characterize
their form and dynamics, and to show that they constitute a basis of the
system Hilbert space.

Several strategies have been proposed to determine the pointer basis. Sorting
all pure states according to their linear entropy production rate allows one
to select the states with minimal loss of purity {\cite{zurek1993a}}. Another
idea considers the solitonic solutions of a nonlinear equation related to the
open quantum dynamics {\cite{Diosi1,Gisin1,Diosi2000a,Strunz2002a}}. Both
approaches were shown to be largely equivalent if a particle is linearly
coupled to a bath of harmonic oscillators {\cite{Diosi2000a}}. The pointer
states are Gaussian wave packets in these linear models
{\cite{Diosi1,Gisin1,Diosi2000a,Strunz2002a}}, and the complete, finite time
decoherence in the Gaussian basis can be proved rigorously for a free particle
{\cite{Eisert2004b}}.

The linear coupling of the system coordinate to the bath assumed in these
studies is often justified on the time scale of dissipation. However, it may
misrepresent the decoherence process since it implies that the decoherence
rate grows above all bounds for increasingly extended quantum states. This
lack of saturation of the decoherence rate is unphysical for local
interactions (an artifact of the unboundedness of the position operator); it
predicts rates, e.g.~for collisional decoherence measurements
{\cite{Hornberger2003afull}}, that are too large by many orders of magnitude.
Moreover, previous studies did not demonstrate the conceptually important
requirement that the statistical weight of an evolving pointer state is given
by its initial overlap with the initial quantum state.

In this paper, we study the emergence of pointer states due to collisional
decoherence, allowing for a realistic and non-perturbative description of the
impact of a gas environment {\cite{Gallis1,kh1}}. We demonstrate how the
orthogonal unraveling of the master equation {\cite{Diosi1986a}} naturally
explains the formation of (non-Gaussian) pointer states and their ensuing
classical evolution, as well as the appearance of the expected statistical
weights. Finally, by relating the widths of the pointer states to the
microscopically defined localization rate of the master equation we can
estimate the coherence length in an interacting thermal gas.

\section{The general pointer basis} To define the notion of pointer
states more precisely, consider a quantum system described by a Lindblad
master equation $\partial_t \rho =\mathcal{L} \left( \rho \right) \,$ due to
its contact with an environment. One may say that $\mathcal{L}$ generates a
pointer basis if the dynamics exhibits a separation of time scales,
characterized by a decoherence time $t_{\ensuremath{\operatorname{dec}}} \,$,
such that there exists a unique set of pure states $\mathsf{P}^{}_{\alpha} = |
\pi_{\alpha} \rangle \langle \pi_{\alpha} |$, which are independent of
$\rho_0$ and which have the property that at all later times any initial state $\rho_0$ can be represented by
a mixture of these states
\begin{eqnarray}
  e^{\mathcal{L}t} \rho_0 & \simeq & _{} \int \mathrm{d} \alpha \, _{}
  \text{Prob}(\alpha|\rho_0)
  \mathsf{P}_{\alpha} (t), \hspace{2em} \mbox{if $t \gg
  t_{\rm dec}$} \, .  \label{eq:decoherence}
\end{eqnarray}
Crucially for interpreting the $\mathsf{P}^{}_{\alpha}$ as proper pointer
states, whenever $\rho_0$
is a superposition of mutually orthogonal pointer states $\mathsf{P}_{\beta}$, 
the ensuing probability distribution $\text{Prob}(\alpha|\rho_0)=\sum_\beta w_\beta\, \delta(\alpha-\beta)$ is determined by the {\em initial} overlaps \begin{eqnarray}
w_\beta=\text{Tr}(\rho_0\mathsf{P}_{\beta}(0))\,.
\label{eq:two}
\end{eqnarray}
$\mathsf{P}_{\alpha}$ evolve in time, slowly compared to the
decoherence scale $t_{{\rm dec}}$, and they initially form a basis (usually overcomplete) with
$\mathrm{d} \alpha$ being the associated measure, $\int \mathrm{d} \alpha \, _{} 
\mathsf{P}_{\alpha} = \mathsf{I}$. 
The uniqueness of the evolving pointer states $\mathsf{P}_{\alpha}$ is a result of equation~(\ref{eq:decoherence}) applying to times $t$, which 
(while exceeding the decoherence time) can be much shorter than the above-mentioned time scale of dissipation, $t_{{\rm dec}}\ll t\ll t_{{\rm diss}}$.

Below, we demonstrate that a viable way of obtaining the $\mathsf{P}_{\alpha}$
and their time evolution is to follow the approach suggested in
{\cite{Diosi1,Gisin1,Diosi2000a,Strunz2002a}} of using the solitonic solutions of
a nonlinear dynamic equation in the space of pure states,
\begin{eqnarray}
  \partial_t \mathsf{P} & = & \left[ \mathsf{P}, \left[ \mathsf{P},
  \mathcal{L} \left( \mathsf{P} \right)]] \: . \label{eqNG}  \label{eq:NG}
  \right. \right.
\end{eqnarray}
This equation is satisfied by the pure state solutions of $\partial_t \rho
=\mathcal{L} \left( \rho \right)$, if there are any, while its nonlinear
character is essential to distinguish pointer states from their
superpositions. It has a natural interpretation as the deterministic part of a
specific class of unravelings of the master equation, as explained below.

Having a solution $\mathsf{P}_t$ to (\ref{eq:NG}) at hand, one may generate
further solutions if the system has an underlying symmetry. In particular, if
$\mathcal{L}$ has the form $\mathcal{L} \left( \rho \right) = 1 / \left( i
\hbar \right) \left[ \mathsf{H}, \rho \right] + \mathcal{D} \left( \rho
\right) \,$ and $\mathsf{U}_t$ is a family of unitary operators satisfying
\begin{eqnarray}
   \mathsf{U} \, \mathcal{D} \left( \rho \right) \,
  \mathsf{U}^{\dag} & = & \mathcal{D} \left( \mathsf{U} \rho \mathsf{U}^{\dag}
  \right) \,,  \label{eq:condition1}\\
   i \hbar \partial_t \, \mathsf{U} & = & \left[ \mathsf{H},
  \mathsf{U} \right],  \label{eq:condition2}
\end{eqnarray}
then $\mathsf{U}_t \mathsf{P}_t \mathsf{U}_t^{\dag}$ can be shown to solve
(\ref{eq:NG}). This will allow us to prove that the pointer states form a
basis, rendering a single pointer state sufficient to construct the whole set.
\begin{figure}[tb]
\centerline{\resizebox{10cm}{!}{\epsfig{file=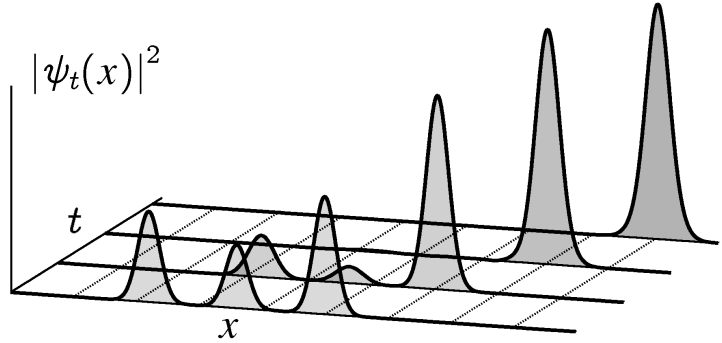}}}
  \caption{The nonlinear equation (\ref{eq:NGCD}) drives any initial state $|
  \psi_0 \rangle$ (here a superposition of wave packets travelling to the
  right) into a localized soliton $| \pi_t \rangle$ that moves with fixed
  envelope and constant velocity. These solitonic solutions form an
  overcomplete set, the pointer basis of collisional
  decoherence.\label{fig:soliton}}
\end{figure}

\section{Collisional decoherence} We now apply these ideas to the 1D
motion of a free particle subject to collisional decoherence
{\cite{Gallis1,kh1,Vacchini2005a}}. The corresponding master equation is the short time version of the full, non-perturbative, Markovian description of a test particle in a gas {\cite{qlbe}}. As such, it describes the loss of coherence but no dissipation effects, and it has the form $\mathcal{L} \left( \rho
\right) = \left( 2 mi \hbar \right)^{- 1} \left[ \mathsf{p}^2, \rho \right] +
\mathcal{D} \left( \rho \right)$, with incoherent part
\begin{eqnarray}
  \mathcal{D} \left( \rho \right) & = & \gamma \int^{\infty}_{- \infty}
  \mathrm{d} q  G \left( q \right) e^{iq \mathsf{x} / \hbar} \rho e^{- iq
  \mathsf{x} / \hbar} - \gamma \rho \, .  \label{eq:CD}
\end{eqnarray}
It involves the collision rate $\gamma$ and the normalized momentum transfer
distribution $G \left( q \right) \,$, each defined by the scattering cross
section and the gas temperature {\cite{Gallis1,kh1}}, and has Lindblad structure
with bounded jump operators $\mathsf{L}_q = \sqrt{\gamma G \left( q \right)}
e^{iq \mathsf{x} / \hbar} \,$. If one disregarded the coherent part, an
exponential decay of the position off-diagonal elements would ensue, $\langle
x| \rho_t |x' \rangle = \exp [- F \left( x - x' \right) t] \langle x| \rho_0
|x' \rangle$\,, with a localization rate $F \left( s \right) = \gamma - \gamma
\int \mathrm{d} q  G \left( q \right) \exp \left( iqs / \hbar \right) \,$,
which saturates at $\gamma$ for $s \rightarrow \infty$.

The form of $\mathcal{L}$ turns the vector representation of (\ref{eq:NG})
into a nonlinear integro-differential equation
\begin{eqnarray}
  \partial_t \psi_t \left( x \right) & = & - \frac{\hbar^{}}{2 mi}
  \partial_x^2 \psi_t \left( x \right) - \psi_t \left( x \right) \Lambda
  \left[ | \psi_t |^2 \right] \left( x \right) \,,  \label{eq:NGCD}
\end{eqnarray}
\begin{eqnarray}
  \Lambda \left[ g \right] \left( x \right) & = & g \ast F \left( x \right) -
  \int^{\infty}_{- \infty} \mathrm{d} yg \left( y \right) g \ast F \left( y
  \right),  \label{eq:lambda}
\end{eqnarray}
involving convolutions $g \ast F \left( x \right) := \int \mathrm{d} y \, g
\left( y \right) F \left( x - y \right)$. Since the dispersion caused by the
first term in (\ref{eq:NGCD}) competes with the localizing effect of the
second term, all solutions of (\ref{eq:NGCD}) turn into stable solitons, $|
\psi_t \rangle \rightarrow | \pi_t \rangle$ for $t \gg
t_{\ensuremath{\operatorname{dec}}}$. These move with constant shape and
velocity, $\left| \pi_{t + \tau} \left( x \right) \right| = \left| \pi_t
\left( x - v_0 \tau \right) \right| \,$, as demonstrated numerically in figure
\ref{fig:soliton}. Below, we show that these solitonic solutions can be
identified with the pointer states of collisional decoherence.

\section{Soliton basis}
Even though the localization rate $F(x)$ remains bounded for large $x$,
the
solitonic solutions to (\ref{eq:NGCD}) are exponentially localized, $\pi_t
\left( x \right)_{} \sim e^{- k \left| x - v_0 t \right|} e^{i \varphi (x -
v_0 t, t)}$ as $x \rightarrow \pm \infty$, with $k > 0$. This follows
analytically by noting that for large $x$ Eq.~(\ref{eq:NGCD}) takes the asymptotic form
\begin{eqnarray}
\partial_t \psi_t \left( x \right) &\sim& - \frac{\hbar}{2 mi}
\partial_x^2 \psi_t \left( x \right) - \left( \gamma - a_{\psi} \right) \psi_t
\left( x \right) ,\quad {\rm as}\; \vert x \vert \rightarrow \infty \, ,
\end{eqnarray}
with a constant $a_{\psi} = \int^{\infty}_{- \infty}
\mathrm{d} y| \psi_t \left( y \right) |^2 | \psi_t |^2 \ast F \left( y \right)
\,$ satisfying $0 < a_{\psi} < \gamma \, \,$. Here, we have used that the convolution
$\vert \psi \vert^2 *\tilde{G}(x)$ vanishes for $\vert x \vert \rightarrow \infty$, where
$\tilde{G}(x)$ denotes the Fourier transform of $G(q)$.
Any solitonic solution $\pi_t
\left( x \right)_{} = f \left( x - v_0 t \right) e^{i \varphi (x - v_0 t, t)}
\,$ thus takes the form $f \left( x \right) = e^{- k \left| x \right|}$ with
$k > 0$ and a phase $\varphi (x, t)$ asymptotically linear in $x$.

Once a particular soliton $\mathsf{P} = | \pi \rangle \langle \pi |$ has been
found, further solitonic solutions to (\ref{eq:NGCD}) are obtained by means of
Galilei transformations. The phase space translation operator $\mathsf{U}_t
\equiv \mathsf{T}_{s, u} = \exp \left( i \left( u_t \mathsf{x} - s_t
\mathsf{p} \right) / \hbar \right)$ satisfies condition (\ref{eq:condition1})
with (\ref{eq:CD}). Moreover, condition (\ref{eq:condition2}) is met with
$\mathsf{H} = \mathsf{p}^2 / \left( 2 m \right)$ provided $s_t = u_0 \, t / m
+ s_0$ and $u_t = u_0$, rendering $\mathsf{T}_{s, u} \mathsf{\, P} \mathsf{\,
T}^{\dag}_{s, u}$ also a solitonic solution of (\ref{eq:NGCD}). This family of
states, parametrized by $\Gamma = \left( s_0, u_0 \right)$, forms an
overcomplete basis since the identity, like any Hilbert-Schmidt operator, can
be represented as $\mathsf{I} = \int \mathrm{d} \Gamma f (\Gamma) 
\mathsf{T}_{s, u} \, \mathsf{Q} \mathsf{\, T}^{\dag}_{s, u} \, \,$ if
$\mathsf{Q}$ is a trace-class operator {\cite{klauder2007a}}.

\section{Orthogonal unraveling} To identify the solitons as pointer
states we now employ the method of quantum trajectories
{\cite{Diosi1986a,unravelkurz,Breuer2007b}}. In this framework, a pure initial
state $\mathsf{P}_0$ is propagated by a stochastic differential equation to
generate an ensemble of pure states $\{ \mathsf{P}^i_t \}_i$, whose average
yields the solution of the master equation, $\mathbbm{E} [ \mathsf{P}^i_t] =
\mathrm{e}^{\mathcal{L} t} \mathsf{P}_0$. Since there are infinitely many
different stochastic processes with this property (unravelings), the ensemble
corresponding to a fixed initial state has no observable consequences, apart form
this average.

However, Eq.~(\ref{eq:decoherence}) is a statement about the solutions for
{{\em all\/}} initial states, such that the physically distinguished basis in
(\ref{eq:decoherence}) may be related to a specific unraveling. In particular,
a preference of a basis is obtained if the deterministic part of the
unraveling exhibits solitonic solutions, such that the stochastic part no
longer affects a trajectory once it has reached these solitons.

We will now show, for the case of collisional decoherence, that this behavior
is found with the {{\em orthogonal unraveling\/}}
{\cite{Diosi1986a}}\footnote{Rather than the original version
{\cite{Diosi1986a}}, we use here the simpler orthogonal unraveling mentioned
in {\cite{rigo1996a}}. It has the advantage that the jump operators are given explicitly.}.
Here, the wave function evolution has a deterministic part given by the nonlinear equation
(\ref{eq:NG}), which is interrupted by random jumps\footnote{We note that Eq.~(\ref{eq:NG}) appears also in the context of a
{{\em diffusive\/}} (i.e. not piecewise-deterministic) unraveling. It was used in {\cite{Diosi2000a}} to study pointer states within a linear coupling model.}. The latter occur with the rate
$r_q = \langle \mathsf{L}_q^{^{\dag}} \mathsf{L}_q \rangle - \langle
\mathsf{L}_q^{^{\dag}} \mathsf{} \rangle \langle \mathsf{}^{} \mathsf{L}_q
\rangle$ and are effected by the (nonlinear) operators $\mathsf{J}_q = \left(
\mathsf{L}_q - \langle \mathsf{L}_q \rangle \right) / \sqrt{r_q}$.

First, we consider only the deterministic part of the quantum trajectory
evolution, given by (\ref{eq:NGCD}). We will restrict the discussion to pure
initial states $| \psi_0 \rangle = \sum_{i = 1}^N c_i | \phi_i \rangle$, which
are superpositions of narrow wave packets, with variance ${\rm Var}_{\phi_i}(\mathsf{x})<2\pi \hbar^2/{\rm Var}_{G}(q)$, 
situated on non-overlapping domains,
$\phi_i \left( x \right) \phi^{\ast}_{j \neq i} \left( x \right) = 0 \,$, with
separation $\Delta x$ sufficiently large such that $F \left( \Delta x \right)
\simeq F \left( \infty \right) \equiv \gamma$. Under this assumption, which will be justified at the end of the following section, 
one can reformulate the nonlinear evolution (\ref{eq:NGCD}) such that the essential
part of the dynamics is already covered by the coefficients $c_i \left( t
\right)$, obeying a closed system of equations,
\begin{eqnarray}
  \frac{\mathrm{d}}{\mathrm{d} t} |c_i |^2 & = & - 2 \gamma  \left( \sum_{j =
  1}^N |c_j |^4 - |c_i |^2 \right) |c_i |^2  \, .  \label{eq:coefficients}
\end{eqnarray}
This equation was studied in \cite{Diosineu} in the context of a model for state vector reduction. As shown there, all its stable fixed points have the form $\left| c_i \right| =
\delta_{i, n}$, and a distinguished fixed point is approached monotonically,  $\left| c_i \left( t \rightarrow \infty
\right) \right| = \delta_{i, m}$, with $m
=\ensuremath{\operatorname{argmax}}_i ( \left| c_i \right|)$ the index of the
the largest wave packet contribution to $| \psi_0 \rangle$.

The dynamics of the wave packets $| \phi_i \rangle$ is described by
\begin{eqnarray}
  \partial_t \phi_i \left( x \right) & = & - \frac{\hbar^{}}{2 mi}
  \partial_x^2 \phi_i \left( x \right) - \phi_i \left( x \right) \Lambda
  \left[ | \phi_i |^2 \right] \left( x \right) \nonumber\\
  &  & + \phi_{^{} i} (x) \sum^N_{j = 1, j \neq i}  \left| c_j \right|^2 
  \tilde{\gamma}_{ij} \left( x \right) \,,  \label{eq:basis}
\end{eqnarray}
with $\Lambda$ defined in (\ref{eq:lambda}) and a rate
\begin{eqnarray}
 \tilde{\gamma}_{ij}(x) &=& \vert \phi_i \vert^2*F(x)-\vert \phi_j \vert^2*F(x)+\gamma \, ,
\end{eqnarray}
which is of the order of $\gamma$. In order to verify (\ref{eq:coefficients}) and (\ref{eq:basis}), one takes
the derivative of $\psi_t (x)=\sum_i c_i (t) \phi_i (t)$ with respect to time; using
\begin{eqnarray}
 \int \mathrm{d}y \vert \phi_i(y)\vert^2 \vert \phi_j\vert^2*F(y) &=& \gamma (1-\delta_{ij})\, , 
\end{eqnarray}
which follows from the above assumptions on $\phi_i(x)$, one finds that
$\psi_t(x)$ evolves according to (\ref{eq:NGCD}). 

Since the coefficients $ c_{j\neq m}$ tend to zero according to (\ref{eq:coefficients}), the coupling term
vanishes after some time in Eq.~(\ref{eq:basis}), turning the latter into the
soliton equation (\ref{eq:NGCD}). In the absence of jumps, the initial
superposition state $| \psi_0 \rangle$ thus evolves into the soliton which is
associated to $| \phi_m \rangle,$ the wave packet contributing largest to $|
\psi_0 \rangle$ (see Fig.~1).

Next, we consider the probabilistic part of the quantum trajectories\,. Using the above
assumption of localized wave packets, $\ensuremath{\operatorname{Var}}_{\phi_i} \left(
\mathsf{x} \right) < 2 \pi \hbar^2 /\ensuremath{\operatorname{Var}}_G \left( q
\right) \,$, one can approximate $\int \mathrm{d} x  e^{iqx / \hbar}
\left| \phi_i \left( x \right) \right|^2 \simeq e^{iqx_i / \hbar}$ with $x_i =
\langle \mathsf{x} \rangle_{\phi_i} \, .$ Under this assumption, the jump
operator $\mathsf{J}_q$ affects only the coefficients. They are transformed as
\begin{eqnarray}
  c_k' \left( q \right) & = & \frac{1}{\mathcal{N}} \left( e^{iqx_k / \hbar} -
  \sum_{j = 1}^N \left| c_j \right|^2 e^{iqx_j / \hbar} \right) c_k \,, 
  \label{eq:jump}
\end{eqnarray}
with $\mathcal{N}$ being the normalization. The jump rate takes the form
\begin{eqnarray}
  r_q & = & \gamma G \left( q \right) \left( 1 - \sum_{j, k = 1}^N \left| c_j
  \right|^2 \left| c_k \right|^2  e^{iq \left( x_j - x_k \right) / \hbar}
  \right) .  \label{eq:jumprate}
\end{eqnarray}
It vanishes for $\left| c_i \right| = \delta_{i, n} \,$, that is, when a
stable fixed point of (\ref{eq:coefficients}) is approached. The associated
wave function can then evolve into a soliton state without being further
perturbed by jumps. This confirms that the quantum trajectories of the
orthogonal unraveling evolve into pointer states, the solitonic solutions to
(\ref{eq:NGCD}).

Concerning the statistical weights, we now show how the stochastic process,
described by (\ref{eq:coefficients}), (\ref{eq:jump}), (\ref{eq:jumprate}) can
be treated analytically for $N = 2 \,$. The average effect of the jump
$\langle c_k' \rangle_G = \int \mathrm{d} qG \left( q \right) c_k' \left( q
\right) \, $ simply interchanges the value of the coefficients, i.e., $|
\langle c_1' \rangle_G |^2 = \left| c_2 \right|^2 \,$. The probability of a
trajectory starting from $\left| c_1 \left( 0 \right) \right|^2 < 1 / 2$ to
converge to $\left| c_i \right| = \delta_{i, 1}$ is therefore equal to the
probability for an odd number of jumps. It is given by $(1 - e^{- 2 \mu \left(
t \right)}) / 2 \,$ for this inhomogeneous Poisson process, with $\mu \left( t
\right) = \int_0^t \mathrm{d} \tau r_{\ensuremath{\operatorname{tot}}} \left(
\tau \right) \, $ the integrated jump rate. In our case,
\begin{eqnarray}
r_{\ensuremath{\operatorname{tot}}} \equiv \int \mathrm{d} q  r_q &=& 2 \gamma
\left| c_1 \right|^2 (1 - \left| c_1 \right|^2) \,,
\end{eqnarray}
where $c_1 \left( t
\right)$ is a solution to (\ref{eq:coefficients}). Rewriting
(\ref{eq:coefficients}) as
\begin{eqnarray}
2 \gamma \left| c_1 \right|^2 (1 - \left| c_1
\right|^2) &=& \frac{1}{2}\partial_t \ln (1 - 2 \left| c_1 \right|^2)  \,, 
\end{eqnarray}
we find that
$\mu \left( t \rightarrow \infty \right) = - \ln (1 - 2 \left| c_1 \left( 0
\right) \right|^2) / 2 \,$. The probability for an odd number of jumps is
therefore $\left| c_1 \left( 0 \right) \right|^2 = | \langle \psi_0 | \phi_1
\rangle |^2$. If $| \phi_i \rangle = | \pi_i \rangle$ this demonstrates that
the statistical weights of the pointer states in the decohered mixture is
given by the overlap $| \langle \psi_0 | \pi_1 \rangle |^2 \,$, in accordance
with (\ref{eq:two}).

The generalization of this result to $N > 2$ can be verified numerically.
Taking $G \left( q \right)$ to be a Gaussian with variance $\sigma_G^2$, we
have implemented the stochastic process (\ref{eq:coefficients}),
(\ref{eq:jump}), (\ref{eq:jumprate}) for various $2 \leqslant N \leqslant
100$, using a Metropolis-Hastings algorithm to draw the momentum transfer $q$
in accordance with the rate (\ref{eq:jumprate}). The initial states $| \psi_0
\rangle$ were generated randomly by simplex picking. A $\chi^2$-test then
confirmed that the asymptotic trajectories, i.e., the pointer states
$\mathsf{P}_i = | \pi_i \rangle \langle \pi_i |$, are distributed according to
$| \langle \psi_0 | \pi_i \rangle |^2 \equiv \ensuremath{\operatorname{Tr}}
\left[ \mathsf{} \rho_0 \mathsf{_{} \mathsf{P}_{}}_i \right]$.

This suggests that the solitonic solutions to (\ref{eq:NGCD}) are the pointer
states of collisional decoherence in the sense that they retain their purity, while
their superpositions decay into mixtures with weights given by the initial overlap. However,
this result was derived under the assumption that the soliton variance is small
compared to  $2\pi \hbar^2/{\rm Var}_{G}(q)$. We will see below that this assumption can be justified for small
$\kappa=\sigma_G^2/m\hbar \gamma$. Still, the finite width leads to a small but finite loss of purity of the solitons,
such that Eq. (\ref{eq:decoherence}) is not an exact equality.

\section{Dynamics and size of the pointer states}
The motion of the
pointer states can be characterized by their position and momentum expectation
values also in presence of an external potential. Our numerical investigations
indicate that the dynamics of this phase space trajectory changes from a
quantum behavior to that expected from the corresponding classical mechanics
for decreasing position spread $\sigma_{\pi}$ of the pointer states. To
explain this fact, we note that an additional linear potential $V \left( x
\right) = ax \,$ in (\ref{eq:NGCD}) implies that the solitonic solutions have
the form $\pi_t \left( x \right)_{} = {f \left( x - x_t \right)} e^{i
\tilde{\varphi} \left( x - x_t, t \right)} \,,^{}$ with $x_t = - at^2 / 2 m +
v_0 t$ and $f$ still exponentially localized. The pointer states thus move
along the accelerated classical trajectories for linear potentials, implying
that they will follow the general classical motion if the linearization of the
potential is permissible over their spatial extension. Similar observations
were made in {\cite{Diosi2000a,bhattacharya2000a}} with linear coupling
models.

The pointer state position spread $\sigma_{\pi} \,$, which depends on the
choice of $G \left( q \right)$ and $\gamma$, thus serves as an important
quantity to characterize the mixture (\ref{eq:decoherence}). We take $G \left(
q \right)$ to be a centered Gaussian with variance $\sigma_G^2$. The
dimensionless form of the soliton equation (\ref{eq:NGCD}) then contains the
ratio $\kappa = \sigma^2_G / m \hbar \gamma$ as the only parameter. By
numerically solving (\ref{eq:NGCD}) we extracted the dimensionless pointer
width $\sigma_{\pi} \sigma_G / \hbar$ as a function of $\kappa$. We found that
the functional form can be reproduced by a simple localization model, which is
inspired by studies of collisional decoherence
{\cite{Hornberger2003afull,Gallis1,kh1}} (and easily extended to the 3D case
below): one assumes that scattering events occurring with rate $\gamma$
localize the wave function to a length scale
$\ell_{\ensuremath{\operatorname{loc}}}^{} \,$, characterized by $1 - F
(\ell_{\ensuremath{\operatorname{loc}}}) / F (\infty) = : \exp \left( -
a_{\ensuremath{\operatorname{loc}}}^2 / 2 \right)$, while it disperses freely
between the collisions. Averaging the wave function width over the waiting
time distribution of a Poissonian process then yields 
\begin{eqnarray}
 \sigma_{\pi} \frac{\sigma_G}{\hbar} &=& \frac{\kappa}{4 a_{{\rm loc}}} + a_{{\rm loc}} \label{eq:size}\, .
\end{eqnarray}
A value of $a_{\ensuremath{\operatorname{loc}}} \simeq 0.4$ reproduces the numerically
obtained widths very well (better than 10\%) over the full range of $\kappa$. It follows 
from (\ref{eq:size}) that for small $\kappa \ll 4a_{{\rm loc}}^2$, the soliton width is given by 
$\sigma_{\pi}^2\simeq a_{{\rm loc}}^2 \hbar^2 / \sigma_G^2$, such that the assumption of
small position variance of $\phi_i(x)$ can be justified for small $\kappa$.
In particular, we have checked numerically the approximation
$\int \mathrm{d} x  e^{iqx / \hbar}
\left| \phi_i \left( x \right) \right|^2 \simeq e^{iqx_i / \hbar}$,
by using the solitonic solution of (\ref{eq:NGCD}); the relative error is less than $2\%$ for $q\in [-2\sigma_G,2\sigma_G]$ and $\kappa \leqslant 10^{-3}$.

\section{Extension to 3D}

Equations~(\ref{eq:decoherence})--(\ref{eq:jumprate}) are
trivially extended to the 3D situation, though their numerical treatment is
then more difficult. However, the above localization model allows one to
directly estimate the 3D pointer width, and thus the coherence length of an
interacting gas, using the microscopically realistic localization rate $F$
derived in {\cite{kh1}}. We illustrate this for the case of hard-sphere
s-wave scattering off a thermal gas, characterized by the thermal wave length
$\Lambda_{\ensuremath{\operatorname{th}}} = \sqrt{2 \pi / mk_{\text{B}} T}
\hbar$. In this case, the localization scale
$\xi_{\ensuremath{\operatorname{loc}}} \equiv
\ell_{\ensuremath{\operatorname{loc}}}^{} /
\Lambda_{\ensuremath{\operatorname{th}}}$ is determined by
$\xi_{\ensuremath{\operatorname{loc}}} = \exp \left(
a_{\ensuremath{\operatorname{loc}}}^2 / 2 - 4 \pi
\xi_{\ensuremath{\operatorname{loc}}}^2 \right) \,
\ensuremath{\operatorname{erfi}} \left( 2 \sqrt{\pi}
\xi_{\ensuremath{\operatorname{loc}}} \right) / 4$, which implies
$\xi_{\ensuremath{\operatorname{loc}}} \simeq 0.1 \,$ if we assume
$a_{\ensuremath{\operatorname{loc}}} = 0.4$, as above. The average over the
waiting time then yields the width of the pointer state
\begin{eqnarray}
  \sigma_{\pi} & = & \, \frac{\ell_{\ensuremath{\operatorname{free}}}}{16
  \xi_{\ensuremath{\operatorname{loc}}} } +
  \xi_{\ensuremath{\operatorname{loc}}}
  \Lambda_{\ensuremath{\operatorname{th}}},  \label{eq:sigmapi3d}
\end{eqnarray}
where $\ell_{\ensuremath{\operatorname{free}}}$ is the mean free path. For a
weakly interacting or thin gas, the pointer state width is thus essentially
determined by $\ell_{\ensuremath{\operatorname{free}}}$. In the limit of a
strongly interacting or dense gas, on the other hand, it is bounded by the
scale of the thermal wave length $\Lambda_{\ensuremath{\operatorname{th}}}$.

We can use (\ref{eq:decoherence}) to characterize the thermal state of a
particle in presence of the interacting gas: the pointer state velocities
display a Maxwell distribution, parametrized by
$\Lambda_{\ensuremath{\operatorname{th}}}$, while their position spread is
given by (\ref{eq:sigmapi3d}). Identifying this state with the reduced single
particle gas state $\rho_{\ensuremath{\operatorname{gas}}}$, we can thus
access the coherence length $\Lambda_{\ensuremath{\operatorname{coh}}} \,$ of
the self-interacting gas. Since the latter is defined by the decay of the
position off-diagonal elements, $\langle \ensuremath{\boldsymbol{x}}|
\rho_{\ensuremath{\operatorname{gas}}} |\ensuremath{\boldsymbol{x}}' \rangle
\propto \exp (- \pi \left|
\ensuremath{\boldsymbol{x}}-\ensuremath{\boldsymbol{x}}' \right|^2 /
\Lambda_{\ensuremath{\operatorname{coh}}}^2) \,$, one obtains
\begin{eqnarray}
  \frac{1}{\Lambda_{\ensuremath{\operatorname{coh}}}^2} & = &
  \frac{1}{\Lambda_{\ensuremath{\operatorname{th}}}^2} + \frac{1}{8 \pi
  \sigma_{\pi}^2}, 
\end{eqnarray}
with $\sigma_{\pi}$ from (\ref{eq:sigmapi3d}) involving the mean free path.
This shows how the interactions in the gas reduce its ideal coherence length,
a behavior possibly observable in the interference of cold, non-degenerate gas
clouds {\cite{Miller2005a}}.

\section{Conclusion} We have shown that the pointer basis of the
collisional decoherence master equation is naturally obtained from the
orthogonal unraveling, explaining the formation and dynamics of the pointer
states, as well as the decay of an initial superposition into the correct
mixture. Since collisional decoherence is a paradigm for the non-perturbative
description of environmental influences, one may expect this unraveling to
generically provide the pointer basis. Apart from its conceptual importance,
it also provides an efficient numerical treatment of open quantum dynamics,
because the quantum trajectories cease to jump after the decoherence time,
leaving the deterministic motion of the pointer states as the only remaining
dynamics.

We thank B. Vacchini for helpful discussions. The work was supported by the
DFG Emmy Noether program.
\vspace*{\baselineskip}


\begin{thebibliography}{10}
  \bibitem{ZurekPointer3mal}W. H. Zurek, Phys. Rev. D {\bfseries{24}}, 1516
  (1981); {\bfseries{26}}, 1862 (1982); Rev. Mod. Phys. {\bfseries{75}}, 715
  (2003).
  
  \bibitem{Joos2003a}E.~Joos {\itshape{et~al.}},
  {\newblock}{\itshape{Decoherence and the Appearance of a Classical World in
  Quantum Theory}} (Springer, Berlin, 2003).
  
  \bibitem{Schlosshauer2007a}M.~Schlosshauer,
  {\newblock}{\itshape{Decoherence and the Quantum-To-Classical Transition}}
  (Springer, Berlin, 2007).
  
  \bibitem{zurek1993a}W.~H. Zurek, S.~Habib, and J.~P. Paz,
  {\newblock}Phys. Rev. Lett. {\bfseries{70}}, 1187 (1993).
  
  \bibitem{Diosi1}L.~Di\'osi, Phys. Lett.~A {\bfseries{122}}, 221
  (1987).

  \bibitem{Gisin1} N.~Gisin and M.~Rigo, {\newblock}J.~Phys.~A: Math. Gen. {\bfseries{28}}, 7375 (1995).
  
  \bibitem{Diosi2000a}L.~Di\'osi and C.~Kiefer, {\newblock}Phys. Rev. Lett.
  {\bfseries{85}}, 3552 (2000).
  
  \bibitem{Strunz2002a}W.~T. Strunz, {\newblock}Lect. Notes Phys.
  {\bfseries{611}}, 199 (2002).
  
  \bibitem{Eisert2004b}J.~Eisert, {\newblock}Phys. Rev. Lett.
  {\bfseries{92}}, 210401 (2004).
  
  \bibitem{Hornberger2003afull}K.~Hornberger, S. Uttenthaler, B. Brezger,
  L. Hackerm\"uller, M. Arndt, and A. Zeilinger, {\newblock}Phys. Rev. Lett.
  {\bfseries{90}}, 160401 (2003).
  
  \bibitem{Gallis1}M.~R. Gallis and G.~N. Fleming, {\newblock}Phys.
  Rev.~A {\bfseries{42}}, 38 (1990).

  \bibitem{kh1} K.~Hornberger and J.~E. Sipe,
  {\newblock}Phys. Rev.~A {\bfseries{68}}, 012105 (2003).
  
  \bibitem{Diosi1986a}L.~Di\'osi, {\newblock}Phys. Lett.
  {\bfseries{114A}}, 451 (1986).
  
  \bibitem{Vacchini2005a}B.~Vacchini, {\newblock}Phys. Rev. Lett.
  {\bfseries{95}}, 230402 (2005).
  
  \bibitem{qlbe}K.~Hornberger, Phys. Rev. Lett. {\bfseries{97}}, 060601
  (2006); K.~Hornberger and B.~Vacchini, Phys. Rev.~A {\bfseries{77}}, 022112
  (2008).
  
  \bibitem{klauder2007a}J.~Klauder and B.~Skagerstam,
  {\newblock}J.~Phys.~A: Math. Gen. {\bfseries{40}}, 2093 (2007).
  
  \bibitem{unravelkurz}C.~W. Gardiner {\itshape{et al.}}, {\newblock}Phys.
  Rev.~A {\bfseries{46}}, 4363 (1992); H.~Carmichael, {\newblock}{\itshape{An
  Open Systems Approach to Quantum Optics}} (Springer, Berlin, 1993);
  K.~M{\o}lmer {\itshape{et al.}}, {\newblock}J.~Opt. Soc. Am.~B {\bfseries{10}},
  524 (1993).
  
  \bibitem{Breuer2007b}H.-P. Breuer and F.~Petruccione,
  {\newblock}{\itshape{The Theory of Open Quantum Systems}} (Oxford University
  Press, 2007).

  \bibitem{Diosineu}L.~Di\'osi, {\newblock}J. Phys. A
  {\bfseries{21}}, 2885 (1988).
  
  \bibitem{rigo1996a}M.~Rigo and N.~Gisin, {\newblock}Quantum Semiclass.
  Opt. {\bfseries{8}}, 255 (1996).
  
  \bibitem{bhattacharya2000a}T.~Bhattacharya, S.~Habib, and K.~Jacobs,
  {\newblock}Phys. Rev. Lett. {\bfseries{85}}, 4852 (2000).
  
  \bibitem{Miller2005a}D.~E. Miller {\itshape{et~al.}}, {\newblock}Phys.
  Rev.~A {\bfseries{71}}, 043615 (2005).
\end{thebibliography}
\end{document}